\documentclass[twocolumn,floatfix,nofootinbib,prd,preprintnumbers,aps,10pt]{revtex4-2}
\pdfoutput=1 
\usepackage[mathscr]{euscript}
\usepackage{amsmath}
\usepackage{graphicx}
\usepackage{dcolumn}
\usepackage{bm}
\usepackage{epsfig}
\usepackage{amssymb,latexsym,mathrsfs}
\usepackage{graphicx}
\usepackage{color}
\usepackage{hyperref}
\usepackage{float}
\usepackage{diagbox}

\usepackage{amsmath}
\usepackage{amssymb}
\usepackage{subfigure}
\usepackage{hyperref}
\usepackage{url}
\usepackage{xcolor}
\usepackage{color}
\definecolor{amaranth}{rgb}{0.9, 0.17, 0.31}
\definecolor{purple(munsell)}{rgb}{0.62, 0.0, 0.77}
\definecolor{americanrose}{rgb}{1.0, 0.01, 0.24}
\definecolor{palatinateblue}{rgb}{0.15, 0.23, 0.89}
\definecolor{royalbluecs}{rgb}{0.2, 0.2, 1}
\definecolor{hanpurple}{rgb}{0.32, 0.09, 0.98}
\definecolor{beaublue}{rgb}{0.74, 0.83, 0.9}
\definecolor{carminered}{rgb}{1.0, 0.0, 0.22}
\definecolor{brightpink}{rgb}{1.0, 0.0, 0.5}
\definecolor{vividviolet}{rgb}{0.62, 0.0, 1.0}
\hypersetup{ linktoc=all, bookmarksopen=true,
    colorlinks=true, linkcolor={royalbluecs},
    citecolor={royalbluecs}, urlcolor={royalbluecs}}
	
\usepackage[open,openlevel=2]{bookmark}

\usepackage{tikz}  

\definecolor{lime}{HTML}{A6CE39}
\DeclareRobustCommand{\orcidicon}{%
	\begin{tikzpicture}
	\draw[lime, fill=lime] (0,0) 
	circle [radius=0.16] 
	node[white] {{\fontfamily{qag}\selectfont \tiny ID}};
	\draw[white, fill=white] (-0.0625,0.095) 
	circle [radius=0.007];
	\end{tikzpicture}
	\hspace{-2mm}
}

\foreach \x in {A, ..., Z}{%
	\expandafter\xdef\csname orcid\x\endcsname{\noexpand\href{https://orcid.org/\csname orcidauthor\x\endcsname}{\noexpand\orcidicon}}
}

\normalsize
\begin{document}

\title{Tests For Maximum Force and Maximum Power}

\author{Christoph Schiller \orcidA{}}
\email{fb@motionmountain.net}
\affiliation{Motion Mountain Research, 81827 Munich, Germany}

\begin{abstract} 
\noindent Two ways to deduce the equivalence of the field equations of general relativity and the principle of maximum force $c^4/4G$ -- or the equivalent maximum power $c^5/4G$ -- are presented. A simple deduction of inverse square gravity directly from maximum force arises. Recent apparent counter-arguments are refuted. New tests of the principle in astronomy, cosmology, electrodynamics, numerical gravitation and quantum gravity are proposed.
\end{abstract} 


\date{16 December 2021} 

\maketitle



\section{Introduction}

\noindent Special relativity is based on an invariant maximum speed $c$ valid for all physical systems.
It is less known that  general relativity can be based on a maximum invariant force valid for all physical systems,  given by 
\begin{equation}
     F_{\rm max}=\frac{c^4}{4G} \approx 3.0 \cdot 10^{43} \rm \,N \;\;. 
	 \label{eq1}
\end{equation}
In the following, two arguments show that the field equations of general relativity follow from maximum force, and vice versa, that the maximum force value follows from the field equations. Maximum force helps in getting an overview of the features and effects of gravity, including the inverse square law, curvature, horizons, black holes and gravitational waves. Recent criticisms of maximum force and maximum power are addressed. Above all, several possible tests in experimental and theoretical research fields are presented. 
Finally, the limits are placed in a wider context that spans all of fundamental physics.

\section{History and experiments}

\noindent The first person to mention maximum force in writing was Rauscher,
in 1973 \cite{rauscher}.  She was followed by Treder \cite{treder}, Heaston
\cite{10.2307/24530850}, de Sabbata and Sivaram \cite{sab} and others
\cite{massa,Kostro:1999ue}.  When the topic was explored in more detail, the
factor 1/4, which is the force limit in natural units, was deduced by Gibbons
\cite{gibbons} and others \cite{csmax}, and studied further
\cite{barrow1,ong2,michael2,barrow2,barrow3,csmax2,Ong:2018xna,b1}.

Maximum force is a consequence of the definition $F=ma$. In relativity, the acceleration of (the front of) a body of length $l$ is known to be limited by $a\leq c^2/l$ \cite{Taylor1983LimitationOP}. As a result, the force on a body of mass $m$ and length $l$ is limited by $F\leq c^2 (m/l)$. The largest ratio $m/l$ arises for a black hole, with a value $c^2/4G$. This yields a maximum force value
$     F_{\rm max}={c^4}/{4G}$, independently of the mass and the length of the body.

Force is also energy per length: a force acting along a path deposes an energy along its length. 
The highest energy per length ratio is achieved when a Schwarzschild black hole of energy $Mc^2$ is deposed over a length given by its diameter $4GM/c^2$.  
This again yields a maximum force of $c^4/4G$. 

Another derivation of the limit arises when considering the force produced by a Schwarzschild black hole on a test mass. When a mass $m$ is lowered, using a string, towards the horizon of a Schwarzschild black hole, the force of gravity $F$  at a radial distance $r$  -- for a vanishing cosmological constant -- is known to be given \cite{ohanian,haye}, to first order, by    
\begin{equation}
    F= \frac{GMm}{r^2\sqrt{1-\frac{2GM}{rc^2}}} \;\;.
    \label{fgfgfgf}
\end{equation}
At first sight,
the expression 
diverges when the test mass approaches the horizon, and thus seems to contradict maximum force.
However, every test mass $m$ is extended in space. To generate a measurable force, the whole test mass needs to be located outside of the horizon.
The test mass itself has a minimum size given by its own Schwarzschild radius ${2Gm}/{c^2}$.
Neglecting spacetime effects due to the test mass by assuming $m\ll M$, 
the minimum size 
yields a smallest possible value for the distance between the centers of both masses. This minimum distance is given by $r=2G(m+M)/c^2$.
Inserting this distance -- which is slightly larger than the black hole radius -- the force of gravitation on the test mass $m$ obeys
\begin{equation}
    F= \frac{c^4}{4G}\, \frac{M\sqrt{m}}{(M+m)^{3/2}} 
    \leqslant \frac{c^4}{4G} \;\;.
\end{equation}
In other terms, the force of gravity felt by a test mass never exceeds the maximum force. This upper limit remains valid  if force is calculated to second order using the results of LaHaye and Poisson \cite{haye}.

Physically, a maximum force $c^4/4G$ is equivalent to a maximum power, or a maximum luminosity, given by 
\begin{equation}
     P_{\rm max}= c \, F_{\rm max} = \frac{c^5}{4G} \approx 9.1 \cdot 10^{51} \rm \,W \;\;, 
	 \label{eqp1}
\end{equation}
corresponding to about $50\,700$ solar masses per second. For comparison, the most  massive known star, R136a1, has about 315 solar masses, whereas black holes, such as the ones in TON68 or in Holm15A, can be as massive as $4$ to $6 \cdot 10^{10}$ solar masses. 

The first to investigate maximum power seems to have been Sciama, also in 1973 \cite{sciama1,sciama2}. 
Others followed \cite{mtw,hogan,Kostro:2000gw,Cardoso:2018nkg,Ong:2018xna,abo,Gurzadyan:2021hgh}. 
The factor 1/4 -- again specifying maximum power in natural units -- arose together with maximum force. 

The maximum force and maximum power values are not well-known. First of all, both values are so large that they do not arise in everyday life, nor under the most extreme experimental situations. In fact, both limit values are only relevant in strong gravitational regimes near black hole event horizons and thus hard to reach.  Secondly, maximum force was only deduced several decades after the development of general relativity, so that it is not found in textbooks. Thirdly, many people are hesitant to use `force' in general relativity. However, force, with its usual definition as change of momentum, $F=dp/dt$, can be freely used also in general relativity. Finally, maximum force leads to several apparent counter-arguments. They are discussed below.


Experimentally, no force value close to the maximum force has ever 
been 
measured. 
The literature is silent on this topic, including the canonical overview of general relativity tests by Will \cite{will}. 
However, in the last few years, checks for the maximum power value are in sight. 

The most powerful known energy sources in the universe are black hole mergers. 
So far, the most powerful events detected by the LIGO and Virgo facilities have reached an instantaneous power of $0.46\pm0.16\,\%$ of the maximum value, namely $230\pm80$ solar masses per second \cite{ligo1}. 
The well-known 2019 black hole merger radiated up to $207\pm50$ solar masses per second \cite{ligo2}. 
Thus, observations with gravitational waves (and simulations) are just 2 orders of magnitude away from potential experimental falsification. 
Future space-based detectors will do better.

Also the luminosity of the full universe did not and does not exceed the value $c^5/4G$. This can be tested in more detail in the future, as shown below.

\section{A short derivation of the field equations}

\noindent Observations during solar eclipses, the constancy of the speed of light, and also the force increase given by expression (\ref{fgfgfgf}) imply that space is \textit{curved} around a mass. %
For example, only taking curvature into account can expression (\ref{fgfgfgf}) be deduced with the dust ball method of Baez and Bunn \cite{baezbunn}. 
In short, maximum force implies that vacuum bends and is elastic.

The elasticity of a material can be described with the \textit{shear modulus}. 
The shear modulus also determines the \textit{shear strength}, i.e., the maximum shear that a material can support (before breaking). 
The two quantities are related by a factor of order $\mathcal{O}(1)$.
Likewise, the elastic constant of the vacuum, $c^4/8 \pi G$, determines, within a factor $\mathcal{O}(1)$, the maximum force $c^4/4G$ that the vacuum can support.

Vacuum elasticity suggests a simple heuristic way to reach the field equations of general relativity starting from maximum force \cite{siva}. 
The energy density $\varepsilon$ in vacuum is a force per area.  A maximum force $c^4/4G$ that also describes the elasticity of vacuum implies
\begin{equation}
	\frac{c^4/4G}{A}  = \varepsilon        \;\;.           
	 \label{eq2}
\end{equation}
This is the maximum energy density for a spherical surface. 
For a spherical surface of radius $r$ and curvature $R=1/r^2$, the area is related to curvature by $A= 4 \pi/R$. 
The relation between curvature $R$ and energy density $\varepsilon$ then becomes
\begin{equation}
	R = \frac{16 \pi G}{c^4} \, \varepsilon         \;\;.           
	 \label{eq3}
\end{equation}
This is the maximum possible curvature for a sphere. 
For a \emph{general} observer, the curvature $R/2$ is replaced by the Einstein tensor $G_{\mu\nu} = R_{\mu\nu} -  g_{\mu\nu}R/2 $,  and the energy density $\varepsilon$ is replaced by the energy–momentum tensor $T_{\mu\nu}$. 
This yields
\begin{equation}
	G_{\mu\nu} =\frac{8 \pi G}{c^4} \, T_{\mu\nu}       \;\;.      
	 \label{eq4}
\end{equation}
This form of the field equations does not yet incorporate the cosmological constant; but it can be extended to do so \cite{siva}. 
In short, using a line of reasoning inspired by vacuum elasticity, the field equations can be intuitively deduced from maximum force.

\section{A longer derivation of the complete field equations}

\noindent Maximum force arises at event horizons. 
Among other properties, all event horizons show energy flow. 
Now, maximum force limits the energy flow through an event horizon. 
This limit allows deriving the field equations.

The simplest finite event horizon is a sphere, characterized by its radius $r$ or, equivalently, by its surface gravity $a=c^2/2r$.  
Event horizons arise from matter or energy in permanent free fall. 
Any falling system at a horizon is characterized by its energy $E$ and its proper length $L$.  
When the fall is perpendicular through the horizon, the momentum change or force measured by an observer at the horizon is given by $dp/dt= F=E/L$.
For a spherical event horizon, the maximum force value and the horizon area $4 \pi r^2$ imply
\begin{equation}
   \frac{E/L}{A} = \frac{ {c^4}/{4G} }{4 \pi r^2} \;\;.
    \label{s5656}
\end{equation}
Horizons being extreme configurations, the left hand side \textit{limits} the amount of energy $E$ of a system with length $L$ flowing through an event horizon of surface $A$.
Now, when a system falls into a horizon, it is accelerated. The geometry of the black hole limits the length $L$ to a maximum value given by the radius
\begin{equation}
    L \leq r = \frac{c^2 }{2a} \;\;.
    \label{LR}
\end{equation}
Combining the last two expressions yields the fundamental relation for every horizon:
\begin{equation}
    E = \frac{c^2}{8 \pi G} \, a\, A \;\;.
    \label{b191203}
\end{equation}
This \emph{horizon equation} relates (and limits) the energy flow $E$ through an area $A$ of a horizon with surface gravity $a$. 
The horizon equation thus follows from and is equivalent to the observation that event horizons are surfaces showing maximum force at every point. 

One notes that the horizon equation also arises if one starts with maximum power instead of maximum force. 
One further notes that the horizon equation is based on test bodies whose speed, acceleration and length are limited by special relativity. 

The next step is to generalize the horizon equation from the static and spherical case to the general case. For a horizon whose curvature varies over space and time, the horizon equation (\ref{b191203}) becomes
\begin{equation}
    \delta E = \frac{c^2}{8\pi G} \, a\, \delta A \quad.
    \label{eq:bb1}
\end{equation}
This differential horizon equation is called the \textit{first law of black hole mechanics} \cite{Bardeen:1973gs,wald}. 
Equating the surface gravity $a$ with temperature and the area $A$ with entropy is a common procedure. 
In this case, the equation is called the \textit{first law of black hole thermodynamics}.  

The first law (\ref{eq:bb1}) describes how a changing horizon area $\delta A$ induces a changing horizon energy $\delta E$ for a given surface gravity $a$.
In other words, the first law  describes the dynamics of every horizon.
In particular, the first law shows that the dynamics of every horizon is determined by the maximum force. 
The situation is analogous to special relativity, where the dynamics for light $x=ct$ is determined by maximum speed.

The first law (\ref{eq:bb1}) is known to be equivalent to general relativity at least since 1995, when this equivalence was shown by Jacobson \cite{jac}. 
The equivalence was confirmed by Padmanabhan \cite{paddy1, paddy2}, by Ashtekar et al. \cite{isohor}, by Hayward \cite{hayward}, and by Oh, Park and Sin \cite{oh}. 
The general argument is the following: using a suitable coordinate transformation, or frame of reference, it is possible to position a horizon at any desired location in space-time. 
This  possibility implies that the dynamics of horizons contains and is equivalent to the dynamics of space-time.
In other words, the first law contains the field equations.

To see in detail how the dynamics of horizons imply the dynamics of space-time, the first law needs to be formulated for arbitrary observers and coordinate systems. 
To achieve this formulation, one introduces the general surface element $d \Sigma$ and the local boost Killing vector field $k$ that generates the horizon (with a suitable norm).  
These two quantities allow rewriting the \textit{left hand side} of the
first law (\ref{eq:bb1}) as
\begin{equation}
    \delta E = \int T_{ab} k^a d\Sigma^b \quad,
\end{equation}
where $T_{ab}$ is the energy-momentum tensor.  
This relation describes horizon energy for an arbitrary coordinates.
	
The \textit{right hand side} of the
first law (\ref{eq:bb1}) can be written
\begin{equation}
    a \; \delta A = {c^2} \int R_{ab} k^a d\Sigma^b \quad,
\end{equation}
where $R_{ab}$ is the Ricci tensor describing space-time curvature.  
This relation describes how the area change of the horizon, given the local acceleration, depends on the local curvature.  
The rewriting \cite{jac,paddy1,paddy2} makes use of the Raychaudhuri equation, which is a purely geometric equation for curved manifolds. 
(The Raychaudhuri equation is comparable to the expression that links the curvature radius of a curve to its second and first derivative.  
In particular, the Raychaudhuri equation does \emph{not} contain any physics of space-time or of gravitation.)

Combining the generalizations of both sides of the first law
(\ref{eq:bb1}) yields the equation
\begin{equation}
    \int T_{ab}k^a d\Sigma^b = \frac{c^4}{8\pi G} \int R_{ab}k^a
    d\Sigma^b \;\;.
    \label{c191203}
\end{equation}
This equation is thus the first law for general coordinate systems and describes the horizon dynamics in the general case. 
Making use of local conservation of energy (i.e., of the vanishing divergence of the energy-momentum tensor), one finds that this equation is only satisfied if
\begin{equation}
   \frac{8\pi G}{c^4} T_{ab} = R_{ab}-\left (\frac{R}{2}+\Lambda \right)
    g_{ab} \;.
    \label{d191203}
\end{equation}
Here, $R=R^c_{\;c}$ is the Ricci scalar. 
The cosmological constant $\Lambda$ arises as an unspecified constant of integration.  
These are Einstein's field equations of general relativity.  

In short, maximum force or maximum power, together with the maximum speed, \textit{imply} the first law of horizon mechanics. 
The first law in turn \textit{implies} the field equations. 
One notes that the derivation only requires the existence of a Riemannian space-time with 3+1 dimensions, and no further conditions.

\section{The principle of maximum force}

\noindent Each step in the previous derivation of the field equations can be reversed: one can return from the field equations (\ref{d191203}) to the first law (\ref{eq:bb1}) and then, using maximum speed, to the maximum force used in equation (\ref{s5656}).

Also the short derivation of the field equations given above using equations (\ref{eq2}) to (\ref{eq4}) can be reversed. 
Again, maximum force arises from the field equations, when maximum speed is taken into account. 

In short, the field equations and maximum force or power are \textit{equivalent}. 
It is therefore acceptable to speak of the \textit{principle} of maximum force or power in general relativity. 
This is akin to speak of the \textit{principle} of maximum speed in special relativity and its equivalence to the Lorentz transformations.

The equivalence of general relativity and of maximum force implies that every test of general relativity \textit{near a horizon} is, at the same time, a test of maximum force. Deviations from general relativity near horizons can be searched for in double pulsars, in black hole mergers, in collisions between neutron stars and black holes, and possibly in other systems \cite{will}. So far, no deviations arose.

\section{Derivation of universal gravity}

\noindent In the \textit{absence} of a horizon, equation (\ref{s5656}) still holds. 
It limits the energy inside a general surface $A$. However, instead of equation (\ref{LR}), special relativity now implies $L \leq 2r = c^2/a$. Equation (\ref{b191203}) then becomes $E= aA \, c^2/4 \pi G$. Inserting $E=Mc^2$ and $A=4 \pi r^2$ results in $a= MG/r^2$.  
Inverse square gravity thus follows from maximum force. 

An even simpler deduction starts with the energy limit per enclosed area 
\begin{equation}
\frac{E}{A}=\frac{F_{\rm max}}{C_{\rm min}} \;\;.
\end{equation}
Then one inserts the area $A=4 \pi r^2$, the maximum force $F_{\rm max}=c^4/4G$ and, from special relativity, energy $E=Mc^2$ and minimum circumference $C_{\rm min}= \pi L_{\rm min} = \pi c^2/a$. 
Together, this yields $a= MG/r^2$, as a direct consequence of maximum force in flat space.

This derivation of the inverse square law does not seem to have been published before.
The lack of the constant $c$ in the inverse square law is thus as natural consequence of the maximum force $c^4/4G$.

\section{Counter-arguments}

\noindent The statement of a maximum force has led to many attempts to exceed the limit.  
First of all, it has to be checked whether Lorentz boosts allow one to exceed the maximum force. 
Since a long time, textbooks show that this is not possible, because both the acceleration and the force values in the proper frame of reference are not exceeded in any other frame \cite{pauli,moeller, Rebhan:1999}. 
(For the simple one-dimensional case, the boosted {acceleration} value is the proper acceleration value divided by $\gamma^3$, while the boosted {force} value is the same as the proper force value.) As a consequence, maximum force is observer-invariant.

What happens if one adds two forces whose sum is larger than the maximum? 
If the forces act at different points, their sum is \textit{not} limited by the principle of maximum force.  
Any force is a momentum flow; the principle does not limit the sum of flows at different locations.
If, instead, the forces in question all act at a single point, the principle states that their sum \textit{cannot} exceed the maximum value. 
In the same way that adding speeds at different points in space can give results that exceed the speed of light, also adding forces at different points in space can give values exceeding the limit.
The speed and force limits are \textit{local.} 
(An incorrect statement on locality is also found in reference \cite{csmax}.)
Recent proposals for exceeding maximum force by Jowsey and Visser \cite{jowsey} explicitly disregarded locality. 
Nevertheless, they were taken up \cite{Faraoni:2021wre}. 
A refutation was first given in reference \cite{newcs} and lead to reference \cite{Faraoni:2021sep}. 
Whenever one tries to exceed maximum force at a specific location, a horizon appears that prevents doing so. 

How can gravitation be the weakest interaction and yet determine the maximum force value? 
Because gravity has only charges of one sign, it is easiest to experience in everyday life. 
However, gravity's ``weakness'' is due to the smallness of typical elementary particle masses, and not to an intrinsic effect \cite{Wilczek:2001ty}. 
In fact, \textit{all} interactions lead to space-time curvature. 
The maximum force value relates curvature to energy density, independently of the type of interaction.

Another potential counter-argument arises from the topic of renormalization of $G$ in quantum field theory. 
The study goes back to the work of Sakharov \cite{Sakharov:1967pk}. 
Various approaches to this issue suggest that $G$ changes with increasing energy, and in particular that $G$ increases when approaching Planck energy. 
This is argued in the papers by Frolov, Fursaev and Zelnikov \cite{Frolov:1996aj},  Visser \cite{Visser:2002ew}, Volovik and Zelnikov \cite{Volovik:2003kt}, and Hamber and Williams \cite{Hamber:2006sv}.  
In contrast, reasons for a fundamental impossibility that $G$ is renormalized were given by Anber and Donoghue \cite{Anber:2011ut,Donoghue:2019clr}.
So far, no hint for a change of $G$ with energy has been found. 
If, however, future experiments do find such such a change, maximum force would be falsified. 

A further potential counter-example is still subject of research. 
Exact calculations on the force between two black holes on the line connecting their centers yield an expression that diverges when horizons touch, thus allowing larger force values at first sight \cite{Krtous:2019uqr}.
However, it appears that those expressions disregard the overall shape changes of the horizons \cite{costaperry}; these shape changes make the horizons touch on a circle around the straight connecting line before they touch on the line. 
Whether this effect prevents exceeding the force limit is still open. 

At least four papers have claimed that the factor in maximum force or power is $1/2$ instead of $1/4$, namely references
\cite{hogan}, \cite{Ong:2018xna}, \cite{abo} and \cite{Gurzadyan:2021hgh}. 
In those papers, the missing factor 1/2 shows up either when distinguishing radius and diameter, or when the factor 2 in the expression $E=2TS$, valid for black hole thermodynamics, is taken into account.

Maximum power has its own paradoxes. 
At first sight, it seems that the maximum power can be exceeded by combining two (or more) separate power sources that add up to a higher power value. 
However, at small distance from the sources, their power values cannot be added. 
And at large distance, the power limit cannot be exceeded, because the sources will partially absorb each other's emission.

A recent theoretical attempt, again by Jowsey and Visser \cite{jow2}, to invalidate the power limit in explosions makes use of an expansion front speed larger than $c$. 
However, the front speed is a signal speed and an energy speed; such speeds are never larger than $c$. 
Equation (\ref{eqp1}) and maximum power remain valid.

In short, no confirmed counter-example to maximum force or maximum power has yet been found. 

\section{Further gravitational limits}
\label{furlim}

\noindent The limits $c^5/4G$ and $c^4/4G$ are not the only ones in general relativity.
An equivalent bound limits mass flow rate  by $dm/dt=c^3/4G \approx 1.0 \cdot 10^{35} \, \rm kg/s $: nature does not allow transporting more mass per time. 
Again, this is a local limit, valid at each point in space-time. 
And again, the limit is realized only by horizons. 
For example, the maximum mass flow rate value limits the speed of a Schwarzschild black hole to the speed $c$.
Again, boosts do not allow exceeding the limit.

The maximum mass rate limit $c^3/4G$ suggests the possibility of future tests, both during the merger of black holes and in numerical simulations.
However, no dedicated studies seem to have been published yet. 

Maximum force also limits mass to length ratios by $c^2/4G \approx 3.4 \cdot 10^{26} \, \rm kg/m $. 
Again, this limit is realized by horizons of Schwarzschild black holes. 
The limit states that for a given mass, nothing is denser than a black hole.
Also this limit cannot be exceeded by a boost: spherical objects, including Schwarzschild black holes, do not Lorentz-contract.
The maximum force thus appears to include the hoop conjecture. 
Again, any counter-example would invalidate maximum force.

In cosmology, more limits arise. Maximum power implies a maximum energy density for the universe. 
Integrating the maximum power $c^5/4G$ over the age $t_0$ of the universe and dividing by half 
the Hubble volume $(2 \pi/3) (c t_0)^3$ yields an upper mass density limit of
\begin{equation}
\varrho_{\rm max} = \frac{3 }{8 \pi G \, (t_0)^2} 
\;\;.
\label{rholim}
\end{equation}
This is the usual critical density. 
In cosmology, the critical density can thus be seen as due to the maximum power $c^5/4G$. 
Indeed, the value is not exceeded in the $\Lambda$CDM cosmological model, nor in measurements. 

In cosmology, expression (\ref{rholim}) for the critical density has further consequences.
Within a factor $\mathcal{O}(1)$, the quantity  $c/4G \approx 1.1\cdot 10^{18}\,\rm kg\, s/m^2 $ appears to limit the product $\varrho\, R_H T_H$ of matter density, Hubble radius and Hubble time \cite{doi:10.1063/1.3536430}.
Similarly, within a factor $\mathcal{O}(1)$, the quantity $1/4G \approx 3.7\cdot 10^{9}\,\rm kg\, s^2/m^3 $ appears to limit the product $\varrho\, T_H^2$ of matter density and (Hubble) time squared. 
Precision tests are under way.

In short, all limits $c^n/4G$ with $0 \leq n \leq 5$ hold. 
They can be tested further with measurements and with simulations. 

Any one of the six gravitational limits $c^n/4G$ can be seen as fundamental.
This also applies to their inverse values. All these limits are \emph{equivalent}.
As a result, also $4G$ is a limit, even though it is not usually seen as one.
Despite this equivalence, speaking of the smallest 
possible value for the inverse of mass density times time squared -- usually called $4G$ --
is somewhat less incisive than speaking of the maximum force $c^4/4G$ or of the maximum power $c^5/4G$.

\section{Alternative theories of gravity}

\noindent Does the maximum force hold in alternative theories of gravity? 
Because general relativity is equivalent to maximum force, the question leads to additional tests.
Dabrowski and Gohar \cite{abo} have shown that maximum force does not apply in theories with varying constants $G$ and $c$. 
However, even the most recent experiments \cite{Wang:2020bjk,Vijaykumar:2020nzc,Le:2021hej} show no such effect.
Dabrowski and Gohar also argue that, similarly, a running of $G$ with energy would invalidate maximum force. Furthermore they show, as did Atazadeh \cite{atazadeh}, that any volume term in black hole entropy invalidates maximum force. 
Atazadeh also explains that quintessence is likely to invalidate the maximum force limit, and so is Gauss-Bonnet gravity. 
Also, maximum force might not be valid in higher spatial dimensions or in conformal gravity. 

It is unclear whether maximum force is invalidated by modified Newtonian dynamics \cite{Milgrom:2012xw}. 
It is seems that not, but the issue is still a topic of research.

In short, maximum force seems to be closely tied to general relativity -- at least near horizons. 
If an alternative theory of gravity is found to describe systems with high curvature, maximum force will be falsified.

\section{Electromagnetic limits}

\noindent Electric charge is quantized in multiples of the down quark charge $-e/3$. 
Electric field is defined as force per charge. 
As a result, a maximum force and a minimum charge imply maximum values for electric and magnetic fields given by $ E_{\rm max} = 3c^4/4Ge = 5.7\cdot 10^{62}\,\rm V/m$ and $ B_{\rm max} = 3c^3/4Ge = 1.9\cdot 10^{54}\,\rm T$.

Unfortunately, the electromagnetic field limits cannot be tested experimentally: in practice, observed field values are limited by the Schwinger field limit, at which pair production arises. The Schwinger field is many
orders of magnitude lower than the Planck-scale limit. 
For this reason, maximum power is not in reach of electromagnetic sources \cite{lu-kumar}. 
Only sources of gravitational waves can achieve values near the power limit.

Could the force between two \textit{charged} black holes be larger than the maximum force? No; the charge reduces horizon radius, but the force limit for test particles remains valid even if the test particle is charged. 
  Explicit calculations of this configuration have been performed in reference \cite{haye}, and more tests will be possible in the future.

Maximum force also implies a limit on the ratio between the magnetic moment and the angular momentum, as deduced by Barrow and Gibbons \cite{ratiolimit}. 
They showed that the ratio is limited by $\mathcal{O}(1)\sqrt{G}/c$, a purely relativistic limit that does not contain $\hbar$. 
So far, this and all other electromagnetic limits thus allow only \emph{theoretical} tests.

\section{Consequences for quantum gravity}

\noindent Maximum force and power hold independently of quantum theory. 
Therefore, the limits can be combined with quantum theory to produce additional insights. 
For example, general relativity alone does not limit curvature, energy density, or acceleration. 
However, limits for these quantities do arise if quantum theory is included.

Combining the limits on speed $v$, force $F$ and action $W$ using the general relation  $Fvt=W/t$ leads to a limit on time measurements given by
\begin{equation}
  t \geq\sqrt{\frac{4G\hbar}{c^5}} \approx 1.1 \cdot 10^{-43}\,\rm s\;\;,
\end{equation}
i.e., twice the Planck time. Shorter times cannot be measured or observed.
Similarly, for acceleration, the relation $Wa={Fv^3/a}$ leads to the limit   $a \leq\sqrt{{c^7}/{4G\hbar}} \approx 2.8 \cdot 10^{51}\,\rm m/s^2$, or half the Planck acceleration. 
Higher accelerations do not arise in nature.

Using the mixing of space and time yields a limit for length given by $l \geq \sqrt{4G\hbar/c^3} \approx 3.2 \cdot 10^{-35}\,\rm m$, twice the Planck length. 
(It thus seems that the existence of actual points in space, which contradicts a smallest measurable length, should at least be put into question.) 
The minimum length in turn leads to limits on area, volume and curvature.
Similar algebra also allows deducing a limit on mass density given by $\rho \leq c^5/(16G^2\hbar)\approx 3.3 \cdot 10^{95}\, \rm kg/m^3$, and a corresponding limit on energy density. 

The quantum gravity limits just deduced are direct consequences of the three basic limits on speed, force and action.
Because the limits prevent the existence of infinite density, infinite curvature and negligible size, they suggest that singularities are not possible, at least for the case of 3 spatial dimensions discussed here. 
This conclusion rises for time-like, space-like, naked and conical singularities. 
(In more dimensions, the situation might differ \cite{barrow3}.) 
For example, the brightest black holes are those with highest density and thus with smallest possible mass: their mass is half the Planck mass. 
But again, during their evaporation, no power larger than $c^5/4G$ is ever emitted.

Another direct consequence of the three fundamental limits arises from the relation $Fl=W/t$, namely the limit $t\,l \geq 4G\hbar/c^4$.
This yields an uncertainty relation relating clock precision and clock size \cite{Bolotin:2016roy} given by
\begin{equation}
    \Delta t \, \Delta l \geq \frac{\hbar}{c^4/4G} \;\;.
\end{equation}
Various analogous uncertainty relations in quantum gravity can be deduced.

A particle is \textit{elementary} -- thus not composed -- if it is smaller than its own  reduced Compton length $\lambda= \hbar / m c$.
Combining this condition with the limits on force, speed and action yields limits on mass, momentum and energy that are valid only for elementary particles:
$E\leq \sqrt{\hbar c^5 / 4G}$ or half the Planck energy,
$p\leq \sqrt{\hbar c^3 / 4G}$ or half the Planck momentum, and  $m\leq \sqrt{\hbar c / 4G}$ or half the Planck mass (thus the opposite limit of that for black hole mass). 
These well-known limits for elementary particles thus also arise from the limits on speed, force and action. 
And indeed, no higher values have ever been observed -- in cosmic rays or anywhere else. 

Combining the limits of this section with the limit on electric charge leads to limits for charge density and for all other electric quantities.
For example, the limits for acceleration and jerk also apply to charged particles. 
The jerk limit therefore limits the  Abraham-Lorentz-Dirac force \cite{Birnholtz:2013nta}. 
Indeed, the force limit is smaller than the maximum force by a factor given by the fine structure constant and a number of order $\mathcal{O}(1)$. 

Also the emission of radiation by an accelerated mirror can be investigated \cite{Myrzakul:2021bgj, Good:2018ell,Zhakenuly:2021pfm}. 
Inserting the limit on acceleration derived above into the expression for the emitted power $P=\hbar a^2/6 \pi c^2$ yields a value that never larger than the maximum power divided by $6\pi$.

Maximum force, together with the quantum of action $\hbar$, also implies a limit on jerk $j$, given by 
\begin{equation}
	j = a/t \leq c^6/(4G\hbar) \approx 2.6 \cdot 10^{94}\,\rm m/s^3 \;\;.
\end{equation} 
It seems that a jerk limit has not been discussed in the literature yet. 
It is known that in the dynamical Casimir effect, the jerk limit implies a power limit. 
Using the usual expression \cite{Good:2017kjr}, the power limit for the dynamical Casimir effect turns out to be given by $c^5/4G$, as expected. This shows again that Planck-scale limits form a consistent set, independent of the specific physical effect under investigation. 
In particular, the limits appear independently of whether the physical effect  explicitly incorporates gravitation or not.

In short, maximum force allows deducing the limits and uncertainty relations usually explored in quantum gravity, including  uncommon ones. 
No contradictions with experiments or with expectations arise.

\section{Thermodynamic limits}

\noindent This rapid overview of quantum gravity did not cover thermodynamic limits that arise by including the Boltzmann constant $k$. 
In 1929, Szilard \cite{szilard} argued that there is a smallest observable entropy of the order of $k$ in nature.
(With its invariance and limit property, the smallest observable entropy $k$ resembles the smallest observable action $\hbar$.)
Including the Boltzmann constant allows deducing an upper temperature limit $\sqrt{\hbar c^5/(4G k^2)}\approx 7.1 \cdot 10^{31}\,\rm K$ given by half the Planck temperature.

Black hole entropy, being a horizon entropy, is the upper limit for the entropy of a physical system with surface $A$, where the surface is a multiple of the smallest surface $A_{\rm min} = {4G\hbar/c^3}$. The factor 4 in the minimum surface is the same factor 4 appearing in the maximum force occurring at horizons. In turn, the factor 4 in the smallest surface appears in black hole entropy, which also occurs at horizons. In short, the factor $1/4$ in black hole entropy is related to the factor $1/4$ in maximum force. 

The Fulling-Davies-Unruh effect and Hawking radiation can also be deduced and allow additional tests.
For example, even an evaporating black hole in its final moments is never hotter than the temperature limit.

\section{Exceeding and approaching the limits}

\noindent What would happen if maximum force or maximum power would be exceeded? 
Exceeding the force limit would mean the ability to affect systems behind a horizon. 
The issue is akin to the ability to circumvent causality by exceeding the speed of light. 
Both are impossible.

Given that maximum force describes the elastic properties of the vacuum, what happens if one gets \textit{close} to the limit?
Just before a material loses its elastic properties, \textit{defects} arise. Similarly, just before the vacuum loses its elastic properties, defects arise; and vacuum defects are particles. 
Indeed, whenever one approaches maximum force by approaching a horizon, particles arise, e.g., in the form of Hawking or thermal radiation. Exploring the microscopic aspects of maximum force and gravitation is subject of ongoing research in quantum gravity.

\section{Searching for a new effect}

\noindent Given that maximum force or power are \textit{equivalent} to general relativity, one does not expect an effect that is specific to maximum force and that is still unknown. 
Nevertheless, one candidate might exist.

Maximum speed $c$ implies a (purely classical) uncertainty relation between frequency and wavelength in wave phenomena given by $\Delta f\, \Delta \lambda \gtrsim c$. 
Minimum action $\hbar$ implies an uncertainty relation between position and momentum in quantum phenomena given by $\Delta x\, \Delta p \gtrsim \hbar$.
This suggests that  an uncertainty relation might exist between observables related by maximum force or power. 
An example is 
\begin{equation}
    \frac{\Delta E}{\Delta l} \lesssim \frac{c^4}{4G} \;\;.
    \label{gur}
\end{equation}
All known systems, such as a typical rock or the Sun, appear to fulfil the inequality.  
The \textit{gravitational uncertainty relation} (\ref{gur}) -- if valid generally -- implies that length uncertainties cannot be zero, but are limited from below by energy uncertainties. 
As a consequence, a quantum vacuum, with its energy fluctuations, cannot be perfectly smooth and flat. 
For a similar reason, due to quantum effects, black hole geometry cannot be perfectly smooth and classical. 
Vacuum and horizons must be cloudy. 
All this is as expected.

More such {gravitational uncertainty relations} can be derived. 
They allow further tests of maximum force and power.

\section{What if maximum force or power would not exist?}

\noindent The question about non-existence of maximum force can be compared to that about the non-existence of maximum speed $c$. 
In the latter case, special relativity would not be valid, light would not be the fastest moving system, 
and, without a natural  invariant standard, speeds could not be measured.
%
Similarly, if force or power would not be bounded, the field equations would not be valid: curvature and energy-momentum tensors would not be connected. 
Also, 
there would be no way to measure force, power, luminosity, mass rate, or mass to length ratio because no natural, invariant standards for them would exist.

\section{The fundamental status of maximum force} 

\noindent One way to state the above results is the following: general relativity \textit{results} from maximum force -- in the same way that special relativity \textit{results} from maximum speed. 
At first sight, this can seem surprising, because physicists are used to think that $G$ and $c$ are fundamental, but not $c^4/4G$. However, as argued in Section \ref{furlim}, there are various possible choices for the fundamental constant of gravity. In particular, it is also possible to take the constants $c$ and $F_{\rm max}$ as fundamental and think of $c^4/4F_{\rm max}=G $ as a derived constant that appears in inverse square gravity. In fact, if desired, one can even take $1/4G$ as a fundamental maximum value of a suitably defined observable, namely mass density times time squared. These -- and other -- choices are all equally fundamental.

Many arguments about maximum force $c^4/4G$ (or any other of its equivalent limits) and maximum speed $c$
can be extended to the elementary quantum of action $\hbar$.
In all three cases, the limit is invariant, cannot be overcome experimentally, leads to apparent paradoxes (as explored for $\hbar$ in the debate between Bohr and Einstein), and yield a specific description of natural phenomena. 

The three limits can be used to express the Bronshtein cube of physical theories -- introduced in the 1930s \cite{Okun:2001rd} -- even more incisively, by using a 
\emph{limit} at every corner of the cube. 
The three upper limits  $1/4G$, $c$ and $1/\hbar$ respectively define non-relativistic gravity, special relativity and quantum theory. Upper limits from combinations, such as $c^4/4G$, $c/\hbar$ and $1/4G\hbar$, respectively define general relativity, quantum field theory and non-relativistic quantum gravity.
Finally, fully combined upper limits such as $c/4G\hbar$ define relativistic quantum gravity. 
In short, one gets a Bronshtein \emph{limit} cube of theories.
If desired, the inverse Boltzmann constant $1/k$ can be added, thus yielding a \emph{limit hypercube} of physical theories \cite{Okun:2001rd}.

The three (or four) fundamental limits also have conceptual consequences.
Special relativity 
predicts the \textit{lack} of physical systems exceeding the speed limit $c$. 
Likewise, general relativity 
predicts the \textit{lack} of physical systems exceeding the force limit $c^4/4G$ (or any other limit equivalent to it). 
For example, there are \emph{no} objects denser than black holes.
Finally, quantum theory predicts the  \textit{lack} of physical systems below the action limit $\hbar$.
Because maximum force allows defining limits in every domain of nature, it predicts the \emph{lack of any trans-Planckian
effect.} Numerous consistency tests, in addition to the ones above, are possible. 
So far, are all positive.

As a final consequence, all invariant limits -- including $c$, $c^4/4G$, $\hbar$, $1/k$, etc. -- are predicted to hold also in a future \emph{unified theory.}
This prediction will be testable in the future.

\smallskip
\section{Conclusion}

\noindent In summary, the principle of maximum force and the principle of maximum power allow deducing general relativity and inverse square gravity. 
The limits are consistent across physics and are useful for teaching and research. 
Searching for counter-examples leads to new experimental tests in black hole mergers and cosmology, and to new theoretical tests in numerical relativity, electrodynamics, quantum gravity and unification. 
So far, no test failed.

\smallskip
\acknowledgments

\noindent The author thanks Michael Good for an intense and productive exchange and Ofek Birnholtz, Barak Kol, Shahar Hadar, Pavel Krtou\v{s}, Andrei Zelnikov, Grigory Volovik, Eric Poisson, Gary Gibbons, Chandra Sivaram, Arun Kenath, Saverio Pascazio, Britta Bernhard, Isabella Borgogelli Avveduti, Steven Carlip and an anonymous referee for fruitful discussions.







 


%

\end{document}